\documentclass{emulateapj}
\usepackage{apjfonts,times}

\newcommand{\beq}{\begin{equation}}
\newcommand{\eeq}{\end{equation}}

\def\alp{\mbox{$\alpha$}}

\def\farcm{\hbox{$.\mkern-4mu^\prime$}}

\def\earth{\hbox{$\oplus$}}
\def\arcmin{\hbox{$^\prime$}}
\def\arcsec{\hbox{$^{\prime\prime}$}}
\def\solar{\mbox{$_{\normalsize\odot}$}}

\newcommand{\AmS}{{\protect\the\textfont2
  A\kern-.1667em\lower.5ex\hbox{M}\kern-.125emS}}
\newcommand{\lsim}{\ \raise
-2.truept\hbox{\rlap{\hbox{$\sim$}}\raise5.truept\hbox{$<$}\ }}
\newcommand{\gsim}{\ \raise
-2.truept\hbox{\rlap{\hbox{$\sim$}}\raise5.truept\hbox{$>$}\ }}
\newcommand{\simsim}{\ \raise
-2.truept\hbox{\rlap{\hbox{$\sim$}}\raise5.truept\hbox{$\sim$}\ }}

\long\def\symbolfootnote[#1]#2{\begingroup%
\def\thefootnote{\fnsymbol{footnote}}\footnote[#1]{#2}\endgroup}

\slugcomment{Accepted for publication in ApJ}

\shorttitle{New view of the Star-Forming Region LH~95/N~64 in the LMC
with HST/ACS Photometry}
\shortauthors{D. A. Gouliermis et al.}

\begin{document}

\title{Discovery of the Pre-Main Sequence Population of the Stellar 
Association LH~95 in the Large Magellanic Cloud with Hubble Space 
Telescope ACS Observations\altaffilmark{1,}\altaffilmark{2}}

\author{Dimitrios A. Gouliermis\altaffilmark{3}, 
        Thomas Henning\altaffilmark{3},
        Wolfgang Brandner\altaffilmark{3,4},\\ 
        Andrew E. Dolphin\altaffilmark{5}, 
        Michael Rosa\altaffilmark{6,}\altaffilmark{7},
        and
        Bernhard Brandl\altaffilmark{8}
}

\altaffiltext{1}{Based on observations made with the NASA/ESA {\em 
Hubble Space Telescope}, obtained at the Space Telescope Science 
Institute, which is operated by the Association of Universities for 
Research in Astronomy, Inc. under NASA contract NAS 5-26555.}

\altaffiltext{2}{Research supported by the German Research 
Foundation (Deutsche Forschungsgemeinschaft) and the German Aerospace 
Center (Deutsche Zentrum f\"{u}r Luft und Raumfahrt).}

\altaffiltext{3}{Max Planck Institute for Astronomy, K\"{o}nigstuhl 
17, 69117 Heidelberg, Germany}

\altaffiltext{4}{UCLA, Div. of Astronomy, 475 Portola Plaza, Los Angeles, 
CA 90095-1547, USA}

\altaffiltext{5}{Raytheon Corporation, USA}

\altaffiltext{6}{Space Telescope European Coordinating Facility, ESO, 
Karl-Schwarzschild-Str. 2, 85748 Garching, Germany}

\altaffiltext{7}{Affiliated to the Space Telescope Operations 
Division, RSSD, ESA.}

\altaffiltext{8}{Leiden University, Leiden Observatory, Niels Bohrweg 2, 
P.O. Box 9513 2300 RA Leiden, The Netherlands}


\begin{abstract} 

We report the discovery of an extraordinary number of pre-main sequence 
(PMS) stars in the vicinity of the stellar association LH~95 in the Large 
Magellanic Cloud (LMC). Using the {\em Advanced Camera for Surveys} 
on-board the {\em Hubble} Space Telescope in wide-field mode we obtained 
deep high-resolution imaging of the main body of the association and of a 
nearby representative LMC background field. These observations allowed us 
to construct the color-magnitude diagram (CMD) of the association in 
unprecedented detail, and to decontaminate the CMD for the average LMC 
stellar population. The most significant result is the direct detection of 
a substantial population of PMS stars and their clustering properties with 
respect to the distribution of the higher mass members of the association. 
Although LH~95 represents a rather modest star forming region, our 
photometry, with a detection limit $V$~\lsim~28~mag, reveals in its 
vicinity more than 2,500 PMS stars with masses down to $\sim 
0.3$~M{\solar}. Our observations offer, thus, a new perspective of a 
typical LMC association: The stellar content of LH~95 is found to extend 
from bright OB stars to faint red PMS stars, suggesting a fully populated 
Initial Mass Function (IMF) from the massive blue giants down to the 
sub-solar mass regime.

\end{abstract}



\begin{figure*}[t!]
\epsscale{.8} 
\plotone{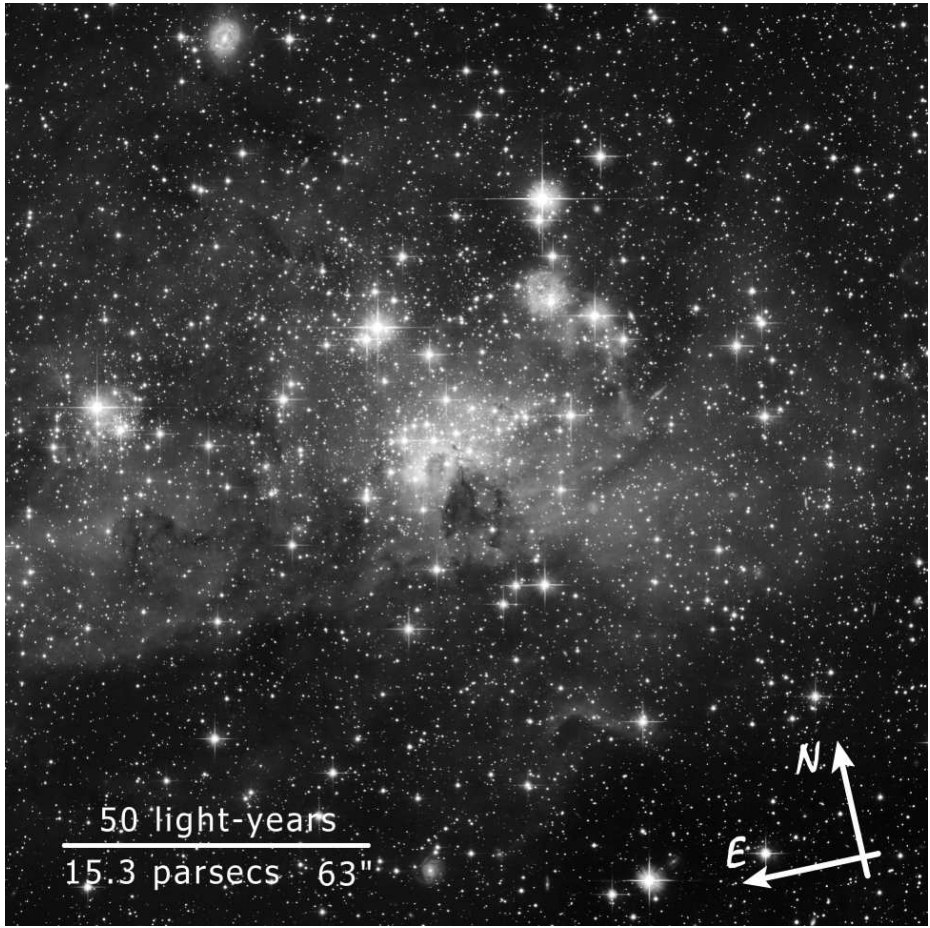}
\caption{Color-composite image from ACS/WFC observations in the filters
$F555W$ and $F814W$ ($V$- and $I$-equivalent) of the LMC star-forming
region LH~95/N~64. This image reveals a large number of low-mass infant
PMS stars coexisting with young massive ones. The smooth continuum in
the image is due to the gas emission from the {\sc H~ii} region, which is
partly included in the $F555W$ filter.\label{fig:pic}}
\end{figure*}

\section{Introduction}
The Large and Small Magellanic Cloud (LMC, SMC) are the closest 
undisrupted neighboring dwarf galaxies to our own. Both the Magellanic 
Clouds (MCs) offer an outstanding variety of young stellar associations, 
the age and IMF of which become very important sources of information on 
their recent star formation. Stellar associations contain the richest 
sample of young bright stars in a galaxy. Consequently our knowledge on 
the young massive stars of the MCs has been collected from photometric and 
spectroscopic studies of young stellar associations (Massey 2006). 
However, a more comprehensive picture of these stellar systems has emerged 
when {\em Hubble} observations revealed that MCs associations host also 
large numbers of faint pre-main sequence (PMS) stars.

Nearby galactic OB associations are known to be significant hosts of PMS
stars (e.g. Preibisch et al. 2002; Brice{\~n}o et al. 2007). The optical
study of significant numbers of such stars in the MCs can only be
achieved at the angular resolution and wide-field facilitated with {\em
Hubble}. Indeed, Gouliermis et al. (2006a) presented, for the first time, 
such a study in the case of the LMC association LH~52, where they
identified the candidate low-mass PMS population of the system with
HST/WFPC2 observations in $F555W$ and $F814W$ filters.  The locations of
the detected PMS stars of LH~52 on the $V-I$, $V$ CMD are found to be in
excellent agreement with the ones of T~Tauri stars with
$M$~\lsim~2~M{\solar} in the Ori OB1 association in the Galaxy (Sherry
et al. 2004; Brice\~{n}o et al. 2005). As far as associations in the SMC
are concerned, Brandner et al. (1999) originally reported the detection
of about 150 objects with excess emission in H{\alp} in a 2\arcmin\
$\times$ 2\arcmin\ field slightly off the brightest stars of the
association NGC~346.  They explained this detection as an indication
that this system hosts PMS stars with masses between 1 and 2 M{\solar}.
More recently, HST/ACS observations allowed Nota et al. (2006) and
Gouliermis et al. (2006b) to verify that indeed there is a prominent
population of low-mass PMS stars located in this association.

In this letter we report the discovery of the PMS stellar content of 
another LMC association, LH~95 (Lucke \& Hodge 1970), related to the {\sc 
H~ii} region LHA 120-N~64, or in short N~64 (Henize 1956). The 
significance of this discovery lies in the fact that our observations, 
being the deepest ever obtained with the {\em Hubble} of a star forming 
stellar system in the MCs, reveal an extraordinary rich sample of faint 
red (with $M$~\gsim~0.3~M{\solar}), as well as more massive (with 
$M$~\lsim~7~M{\solar}) but short-lived PMS stars. In \S 2 we describe the 
data used in this study, and in \S 3 we present the discovery of these 
stars and a typical CMD of an LMC association, as our new {\em Hubble} 
data reveal it. We also discuss the spatial distribution of these stars. 
A short summary and future prospects are given in \S 4.

\section{Observations and Photometry}

The data used in this study were collected within our HST GO Program 
10566. Two pointings (with significant offsets) were observed with the 
Wide-Field Channel (WFC) of the {\em Advanced Camera for Surveys} (ACS) in 
the filters $F555W$ and $F814W$, equivalent to standard $V$ and $I$ 
respectively. The first pointing is centered on the association LH~95. The 
second pointing is located $\sim$~2\arcmin\ to the west, on a nearby 
``empty'' area, selected as the most representative of the local 
background field of the LMC. We refer to these observed fields simply as 
the ``system'' and the ``field'', respectively. A color-composite image of 
the pointing on the system is shown in Figure \ref{fig:pic}. These 
observations, being among the deepest ever taken towards the LMC, allow us 
to explore the scientific gain that can be achieved for the study of 
resolved extragalactic stellar populations using high spatial resolution 
photometry from {\em Hubble}\footnote{Press releases related to this study 
can be found at {\tt www.spacetelescope.org/news/html/heic0607.html} and 
{\tt hubblesite.org/newscenter/archive/releases/2006/55/}.}.

Photometry was obtained using the ACS module of the package 
DOLPHOT\footnote{The ACS mode of {\tt DOLPHOT} is an adaptation of the 
photometry package {\tt HSTphot} (Dolphin 2000). It can be found with its 
documentation at {\tt http://purcell.as.arizona.edu/dolphot/}.} (Ver. 
1.0). We followed the photometric process as it is described by Gouliermis 
et al. (2006b) for similar observations of the association NGC~346 in the 
SMC. One of the advantages of DOLPHOT is the ability to perform photometry 
simultaneously for all exposures. We did this using the $F814W$ drizzled 
frame as the position reference for the detected stars. Photometric 
calibrations and transformations were made according to Sirianni et al. 
(2005), and CTE corrections were made according to ACS ISR 04-06. A 
complete acount of the photometric procedure will be given in a subsequent 
paper. After cleaning bad detections based on DOLPHOT's star quality 
parameters, more than 16,000 stars were detected in the system and 17,000 
in the field. The completeness of the data was evaluated by artificial 
star experiments, that were performed by running DOLPHOT in artificial 
star mode. The completeness, which reach the limit of 50\% at $V 
\approx$~28~mag in the system, as well as the photometric uncertainties, 
which are $\sigma$~\lsim~0.1~mag for $V$~\lsim~27.5~mag, will be also 
discussed in detail in a subsequent paper.

\begin{figure*}[t!]
\centerline{\hbox{
\includegraphics[width=0.33\textwidth,angle=0]{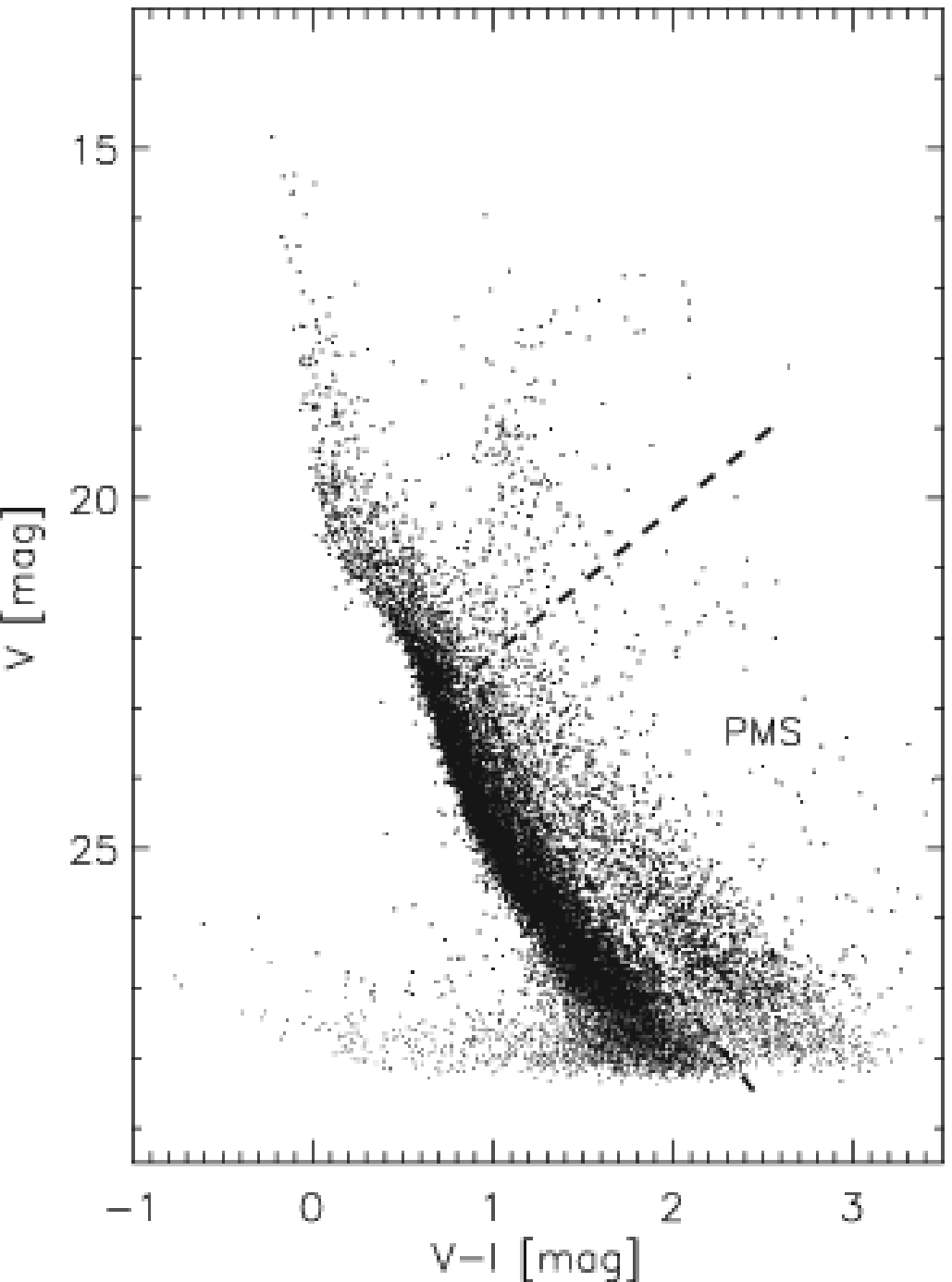}
\includegraphics[width=0.33\textwidth,angle=0]{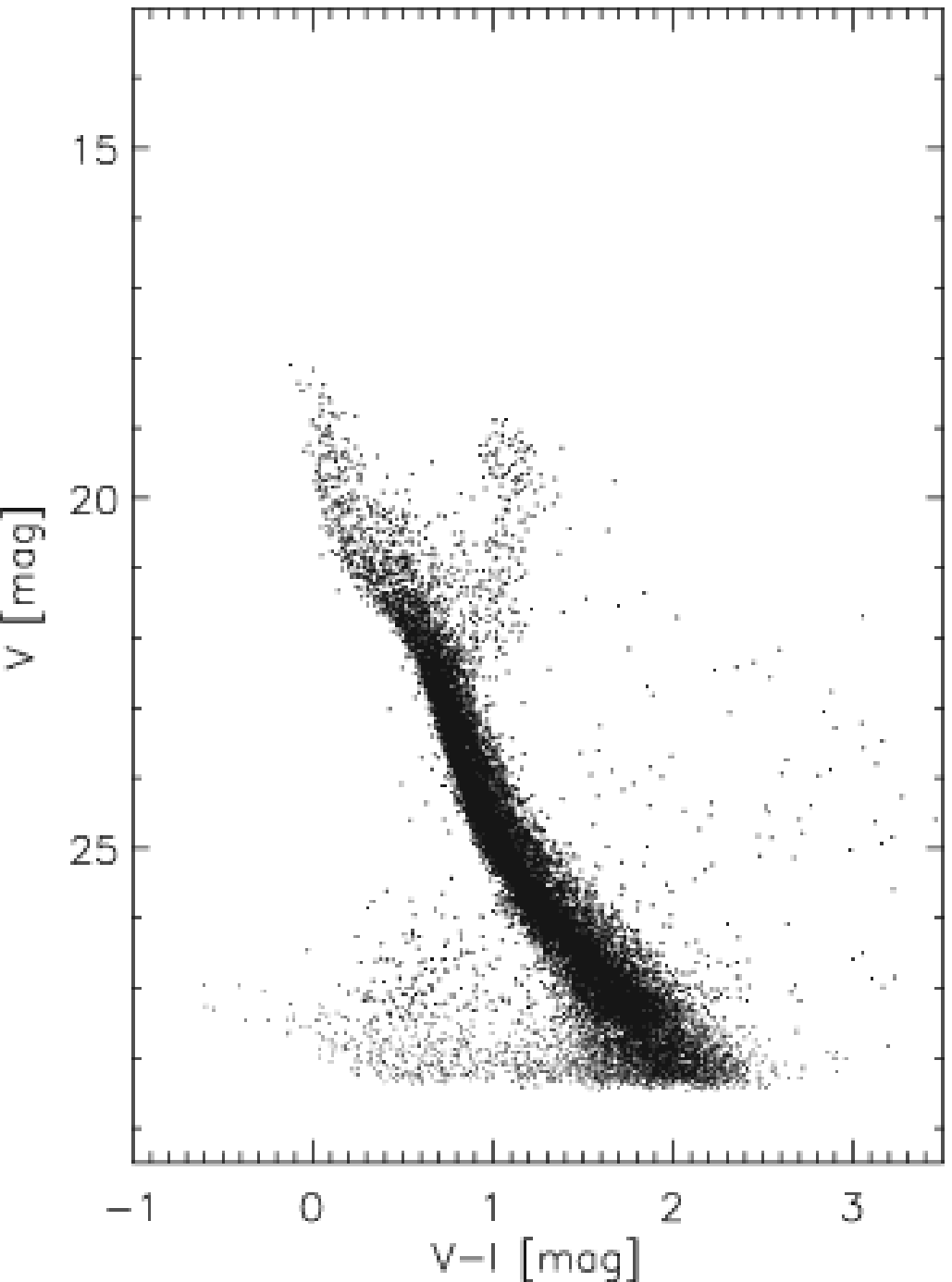}
\includegraphics[width=0.33\textwidth,angle=0]{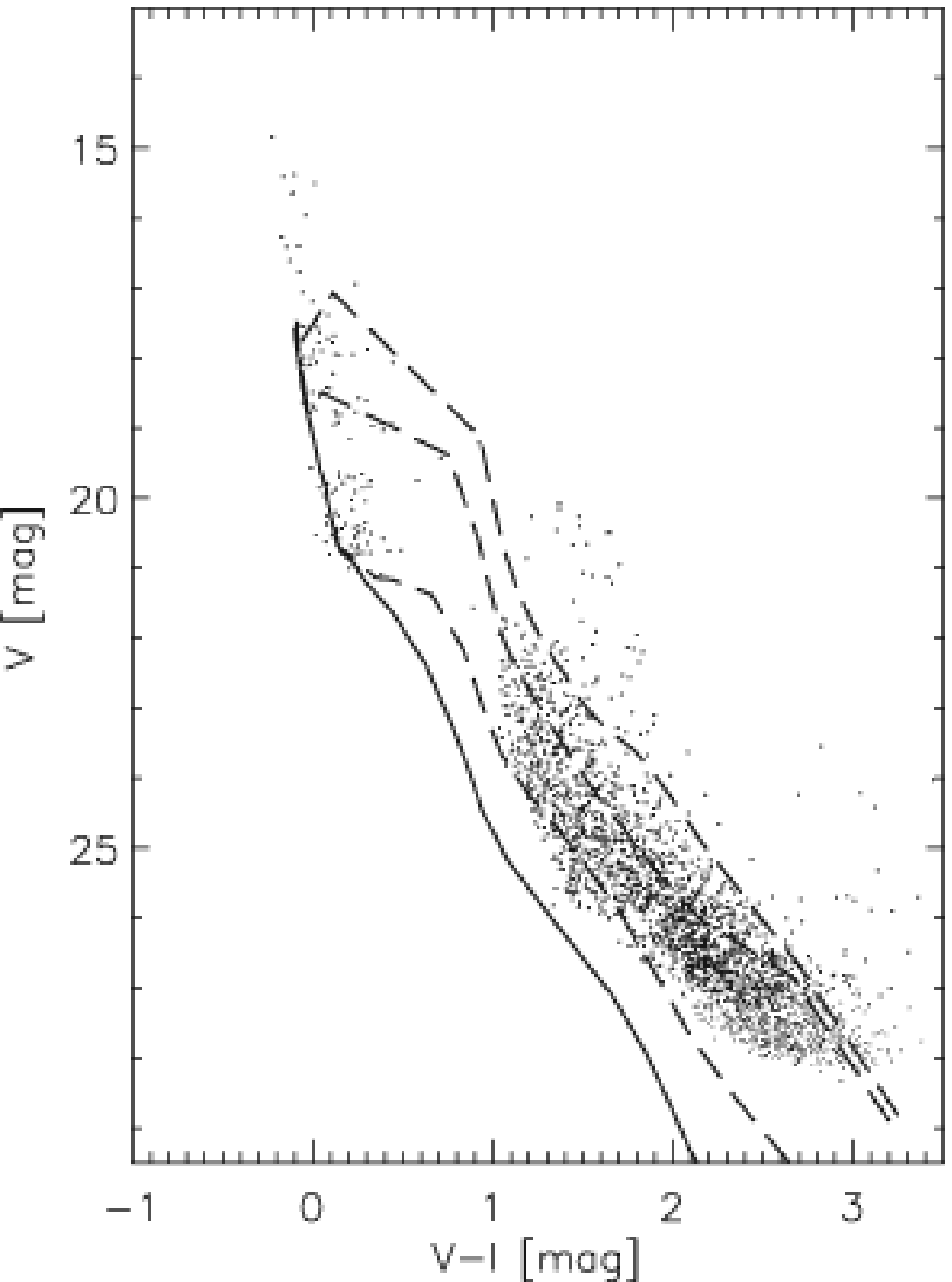}
}}
\caption{The $V-I$, $V$ CMDs of the stars detected with our photometry in 
the system (left) and the field (middle). The comparison of these CMDs 
exhibits the differences in stellar content of the two observed fields, 
and demonstrates the richness of PMS stars in the vicinity of a LMC 
stellar association, as it is emphasized by the complete lack of such 
stars in the LMC field. This is clearly shown in the CMD of the 
association after the contaminating stellar population of the LMC field 
has been statistically subtracted (right). Three PMS isochrones (for ages 
0.5, 1.5 and 10 Myr; dashed lines) as well as the ZAMS (solid line) from 
Siess et al. (2000) grid of models are overplotted. These models suggest 
that PMS stars of masses from $\sim$~7~M{\solar} down to 
$\sim$~0.3~M{\solar} (within 50\% completeness) exist in the association. 
In all CMDs stars with the best photometry ($\sigma \leq$~0.1~mag in both 
bands) are plotted with thick symbols. \label{fig:cmds}}
\end{figure*}

\section{Detection of the Pre-Main Sequence Stars}

\subsection{Color-Magnitude Diagrams}

The $V-I$, $V$ Color-Magnitude Diagram (CMD) of the detected stars in both 
the system and the field is shown in Figure \ref{fig:cmds} ({\em left} and 
{\em middle} panel, respectively). This figure shows that both regions 
provide excellent examples of well resolved extra-galactic stellar 
populations, with the CMDs being characterized by the coexistence of 
different stellar species. In the CMD of the system (Figure \ref{fig:cmds} 
{\em left}) there is a sharp upper main sequence (UMS) of young blue 
massive stars, which is characteristic of LMC associations. There is a 
bright cutoff in the CMD of the field (Figure \ref{fig:cmds} {\em middle}), 
because of the lack of short exposures, but previous ground-based 
photometry by Gouliermis et al. (2002) has shown that this area does not 
include any UMS stars. On the other hand, a pronounced turn-off at around 
$V\simeq 22.5$~mag and a red giant branch (RGB) with its red clump clearly 
located at around $V \simeq 19$~mag and $V-I \simeq 1.2$~mag, typical of 
the LMC field (e.g. Elson et al. 1997; Smecker-Hane et al. 2002) can be 
seen in both CMDs. Below the turn-off, moving to fainter magnitudes, the 
main sequence in both CMDs is increasingly densely populated with what 
appears to be a well mixed collection of low main sequence (LMS) stars. 
However, apart from the UMS, a striking difference between the two CMDs 
exists in the red part of the LMS, where a prominent sequence of PMS stars 
can be seen in the CMD of the system, but is completely absent from the 
field.

A first-order selection of the part of the CMD occupied by these stars,
as it is drawn with a thick dashed line in Figure \ref{fig:cmds} ({\em
left}), reveals the extraordinary number of at least 2,500 PMS stars in
the 3\farcm4~$\times$~3\farcm4 wide area of this rather quiescent star
forming region in the LMC. Moreover, the locations of these PMS stars in
the CMD fit very well the ones of the PMS stars previously discovered in
other MCs associations (e.g. Gouliermis et al. 2006a,b), and they match
the ones of T~Tauri stars as they have been observed over relatively
similar wavelengths in OB associations of the Galaxy (Sherry et al.
2004; Bricen\~{o} et al. 2005). It should be noted that although there
are conspicuous dust clouds within the association (Figure
\ref{fig:pic}), the interstellar reddening in the region is not high
enough to cause the apparent difference at the PMS part of the CMD
between the system and the field.

\subsection{The CMD of the Stellar Association}

Both areas of the system and the field were observed with identical setup 
of the instrument, dithering and exposure times. However, for the system a 
few additional short exposures were taken to handle over-exposed stars in 
the brightest upper part of the CMD. The identical observing procedures for 
system and field allows us a direct decontamination of the system CMD from 
the contribution of the local LMC field, as it is captured in the CMD of 
the field. We perform this process by applying a Monte Carlo subtraction 
method. Specifically, we divided the CMDs of both areas in similar sets of 
subregions, and we statistically subtracted from the CMD of the area of 
the system the corresponding number of randomly selected stars in the CMD 
of the field for each of these subregions. The derived ``clean'' CMD of 
the association alone is shown in Figure \ref{fig:cmds} ({\em right}), where 
it is verified that indeed it is the UMS and PMS populations that define 
such systems. According to PMS isochrone models by Siess et al. (2000) the 
PMS stars of LH~95 cover a mass-range down to 0.3~M{\solar} (within 50\% 
completeness). It should be noted, though, that in this CMD we cannot 
detect the turn-on as it would be expected by the overplotted models. This 
is because this part of the CMD coincides with the turn-off of the field 
population, which is quite rich (as is seen in both {\em left} and {\em 
middle} panels of Figure \ref{fig:cmds}). A more detailed study of the 
contamination of the turn-on by the turn-off will be performed in a 
forthcoming paper.

\begin{figure*}[t!]
\epsscale{1.}
\plotone{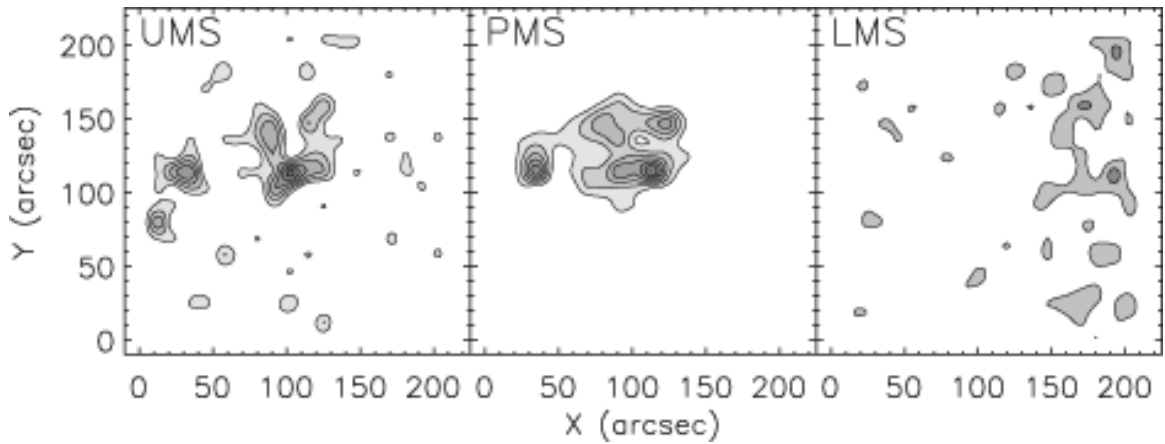}
\caption{Isodensity contour maps constructed from star counts of three
different stellar populations, which cover different regions in the CMDs
of Figure \ref{fig:cmds} (UMS, PMS, and LMS). From these maps it is shown
that PMS stars are forming compact subgroups, which coincide with the
higher concentrations of UMS stars as well. LMS, as well as RGB, stars do not
appear with any statistical significance in the region of the system,
probably due to foreground extinction by the dust and crowding.\label{fig:cm}}
\end{figure*}

\subsection{PMS sub-clusters in Stellar Associations of the LMC}

In order to visualize the spatial distribution of the various stellar 
content within the system we performed star counts on our catalog of 
detected stars in square grids, under the assumption that each star is a 
point determined by its coordinates in the catalog. The binning was done 
so that each grid element has dimensions $\simeq$~216 WFC pixels, or 
$\simeq$ 10\arcsec, which corresponds to about 2.5~pc at the distance of 
the LMC. This technique allows the construction of detailed maps of the 
spatial density distributions for each stellar group (isodensity contour 
maps). The constructed contour maps for the UMS, PMS and LMS are shown in 
Figure \ref{fig:cm}. Surface density is plotted in steps of 1$\sigma$, where 
$\sigma$ is the standard deviation of the background density, starting 
from the 1$\sigma$ isopleth.

Any stellar concentration with density equal or higher than 3$\sigma$ 
above the background density (third isopleth in the maps of Figure 
\ref{fig:cm}) is considered to be statistically significant. One can see 
that both, the UMS and the PMS stars are non-uniformally distributed within 
LH~95. Their spatial distribution is quite clumpy on the scale of 
20\arcsec\ ($\sim$~5~pc), and both species form several density peaks, 
four of which coincide between the UMS and PMS maps. This clearly suggests 
that the PMS stars are concentrated in small clusters, which are 
characterized by bright massive stars.  These PMS clusters do not show to 
be physically independent to each other, but they probably are subgroups 
of the association itself. This is a typical case also for local OB 
associations in our galaxy (Bricen\~{o} et al. 2007). As far as the LMS 
concerns, their isodensity contour map shown in Figure \ref{fig:cm} ({\em 
right}) indicates that these stars do not form any 
significant concentrations in the area of LH~95. This is also the case for 
the RGB stars.

\section{Summary}

We take advantage of the large improvement in sensitivity and wide-field 
resolution provided by ACS to perform a detailed photometric study of the 
star-forming region LH~95/N~64 in the LMC. We performed photometry on two 
WFC pointings, 3\farcm4~$\times$~3\farcm4 each, centered on the 
association LH~95 and a nearby representative background field. Since both 
pointings were observed with similar telescope settings, the derived 
photometric catalogs provide a coherent stellar sample for both observed 
regions, allowing us to remove accurately the contamination of the field 
from the CMD of the system with the use of a Monte Carlo technique. These 
one-of-a-kind observations dramatically enhance the picture we have had up 
until now for stellar associations in the Magellanic Clouds by revealing a 
unique rich sample of more than 2,500 low- and intermediate-mass PMS stars 
(with 7~\lsim~$M$~\lsim~0.3~M{\solar}) in LH~95/N~64. Our data revealed, 
thus, the complete CMD of a stellar association in the LMC from the 
massive regime populated by the brightest UMS stars down to the lowest 
masses ever observed in this galaxy, where PMS stars dominate.

The spatial distribution of the PMS members of the association 
demonstrates the existence of significant substructure (``subgroups''), as 
in the case of galactic OB associations.  This stellar sub-clustering has 
been suggested to have its origins possibly in short-lived parental 
molecular clouds within a Giant Molecular Cloud Complex (Brice\~{n}o et 
al. 2007). Each of these ``{\em PMS clusters}'' in LH~95/N~64 includes a 
few early-type stars. Such stars have been identified as candidate Herbig 
Ae/Be (HAeBe) stars due to their strong H\alp\ emission (Gouliermis et al. 
2002). A more comprehensive study of the star-forming region of LH~95/N~64 
will include the further detailed investigation of the PMS/UMS subgroups, 
which will be performed with the use of our ACS data in synergy with 
near-IR spectroscopy obtained with SINFONI on ESO/VLT.

\acknowledgments

D. A. Gouliermis kindly acknowledges the support of the German Research 
Foundation (Deutsche Forschungsgemeinschaft - DFG) through the individual 
grant 1659/1-1. We also acknowledge the contribution of M. Schmalzl 
through his field-subtraction Monte Carlo algorithm. 



\end{document}